THE EUROPEAN
PHYSICAL JOURNAL C

Regular Article - Theoretical Physics

# Lorentzian wormhole in the framework of loop quantum cosmology

Rikpratik Sengupta[1,a], Shounak Ghosh[2,b], Mehedi Kalam[1,c]

[1] Department of Physics, Aliah University, Kolkata, West Bengal 700160, India
[2] Directorate of Legal Metrology, Department of Consumer Affairs, Government of West Bengal, Malbazar, Jalpaiguri, West Bengal 735221, India

**Abstract** In this paper, we construct a traversable static Lorentzian wormhole in the effective scenario of Loop Quantum Cosmology (LQC), where the field equations are modified due to the ultraviolet (UV) corrections introduced at large space-time curvatures. A stable wormhole can be constructed in the effective scenario without the violation of Null energy condition (NEC) by physical matter at the throat. The NEC is effectively violated due to the corrections in the field equations from LQC, resolving the Weyl curvature singularity at the throat. However, the physical matter does violate the Strong energy condition (SEC), suggesting the interesting possibility that dark energy can be harnessed into a wormhole. A possible explanation for this is the presence of inherent pressure isotropy in the UV-corrected field equations (discussed and compared to braneworld wormholes in the discussion). No additional exotic ingredient (violating NEC) is required, avoiding quantum instabilities. The tidal forces at the throat do not diverge and also the throat is found to be stable. The wormhole features an attractive geometry. LQC can resolve both types of curvature singularities appearing at the black hole centre and wormhole throat, without exotic matter.

## 1 Introduction

Wormholes are geometrical structures which appear as a solution to the field equations of Einstein's General Relativity (GR). Although one of the most attractive predictions of GR, they have not yet been detected directly by observations. Einstein himself along with Rosen [1] visualized wormholes as space-time bridges connecting two different space-time points across the universe, acting as shortcut paths of space-time travel between them. The first rigorous mathematical study of wormholes was performed by Fuller and Wheeler in the early 1960s [2], a decade marking lot of the formal developments of modern GR that changed the outlook towards the subject. However, the result obtained by them left wormholes to be a subject of academic interest only. In fact, the path breaking paper of Morris and Thorne [3] in 1988 that caused a revolution in wormhole physics was originally considered as an academic tool for better understanding of GR, but the profound implications of the results obtained made them write a second follow up paper on the subject in the same year [4].

Fuller and Wheeler had found that although wormhole geometry described by tubular shaped objects with two openings (mouths) spreading out to be asymptotically flat at infinitely large radial distances from the throat (narrow region connecting the two mouths) did exist as static, spherically symmetric solutions to the Einstein field equations (EFE), realistic Schwarzschild wormholes were unstable at the throat due to development of infinitely large gravitational tidal forces resulting in a Weyl curvature singularity at the throat (diverging Weyl tensor). The generation of the large tidal forces can be understood physically from the fact that the matter at the throat be attracted gravitationally by the two mouths in opposite directions. The idea of Morris–Thorne (MT) to avert this Weyl singularity was simple and elegant. If the matter at the throat be replaced by a form of exotic gravitationally repulsive matter, then the tidal forces developing at the throat can be prevented from diverging. However, one has to pay the price that considering the energy density of such matter always remains positive, the Null Energy Condition (NEC) $\rho + p \geq 0$ has to be violated by the matter at the throat. Although, a violation of the Strong energy condition (SEC) $\rho + 3p \geq 0$ is an essential condition to obtain an accelerating universe in a cosmological context and has been

a e-mail: rikpratik.sengupta@gmail.com
b e-mail: shounakphysics@gmail.com
c e-mail: kalam@associates.iucaa.in (corresponding author)

Springer

realized in inflationary and dark energy models involving scalar fields or non-linear equation of states (EoS), violation of the NEC is an even bigger ask and may lead to quantum instabilities of the vacuum. This issue shall be taken up in more details later in this paper.

Another very important factor in wormhole physics is the radial metric potential of the static, spherically symmetric metric known as the shape function, as it determines the shape of the wormhole. As per the MT prescription, in order to build a traversable wormhole that can possibly allow any form of human traversability with limited tidal forces preventing the traveller from getting ripped apart at the throat, the shape function $b(r)$ must satisfy a number of criteria, given as follows: (i) the shape function at the throat radius $r_0$ must be equal to the throat radius itself ($b(r_0) = r_0$). (ii) For radial distances $r > r_0$, the ratio of the shape function at any given radial distance $r$ to that radial distance must be less than unity ($\frac{b(r)}{r} < 1$). (iii) The first derivative of the shape function with respect to the radial distance $r$ at the throat must be less than unity ($\frac{db(r)}{dr}|(r = r_0) < 1$). The final condition implies a minimal throat size, thereby minimizing the amount of exotic matter required at the throat to violate the NEC.

In order to violate any of the energy conditions, either the matter sector or the geometry sector of the EFE has to be modified via a modification in the matter or gravitational Lagrangian. Such modifications can alter the relativistic behaviour either at the ultraviolet (UV) or infrared (IR) scales through correction terms in the EFE. The presently observed acceleration of the universe [5,6] at low energy (IR scale) requires a violation of at least the SEC (some models violate the NEC also). This can be sourced by modifications in the matter sector via – minimally coupled scalar fields dubbed as the quintessence [7,8] with suitable steep potentials, by a fluid known as Chaplygin gas [9,10] that is described by a non-linear EoS and finds its origin in extra dimensional theories, or a phantom fluid that is described by a supernegative EoS with an EoS parameter $< -1$ [11–13]. Alternatively, late time acceleration can also be achieved by modifying the geometry sector [14–16]. At the UV scale it is more useful to modify the geometry sector due to the high energy density and large space-time curvature. The two most acceptable effective modified scenarios in this context are the Loop Quantum Cosmology (LQC) [17,18] and the braneworld scenario [19,20].

Traversable wormholes have been constructed in literature with both the approaches modifying matter [21–32] as well as geometry [33–44] sectors. A possibility of existence of wormholes in certain regions of our galaxy has been explored recently [45]. Both the UV corrected effective scenarios are known to resolve the strong Ricci curvature singularity at the centre of the black hole [46,47]. Also, the initial big bang singularity is found to be resolved in the LQC scenario [17,18]. This is a key motivation behind the attempt to construct traversable wormholes using these UV corrected frameworks, where the Weyl curvature singularity may be resolved at the wormhole throat to make them traversable. We have successfully constructed traversable wormhole in the Randall–Sundrum II (RSII) braneworld scenario [41]. The RSII model has an inherent $Z_2$ symmetry which is absent in LQC. It has been shown by Konoplya and Zhidenko [48] that a fully consistent traversable wormhole can be constructed from normal matter with coupled Maxwell and Dirac fields in the absence of $Z_2$ symmetry at the throat in a relativistic context. The LQC scenario can be realized in $(3 + 1)$-dimensions and one need not be sceptical about the existence of extra dimensions. In this paper, we attempt to construct a traversable wormhole in the framework of LQC which is an effective scenario that avoids the conceptual problems arising from the quantum mechanical interpretations of the gravitational system and helps to provide a much better understanding of the classical singularity. We solve the modified EFE in the LQC scenario for a spherically symmetric matter distribution to obtain the wormhole shape function and also check the validity of the NEC in the effective matter description. The unknown model parameters are estimated by applying the junction conditions at the wormhole surface. The components of the tidal acceleration at the wormhole throat have been computed and confining them to physically justifiable values, an upper limit on the velocity of the traveller traversing the wormhole is obtained. Also, a linearized stability analysis is performed to ensure stability of the traversable wormhole and the nature of the wormhole geometry can be inferred from obtaining the radial acceleration. We conclude with a discussion on the physical consequences of the results obtained.

## 2 Mathematical model of the wormhole

In this section a static, spherically symmetric and traversable wormhole model is constructed that is stable under the linearized stability analysis. The validity of the NEC is checked along with the traversability criteria computing the tidal forces at the throat. The junction conditions have been made use of to determine the unknown model parameters which have been used to make the plots. The surface density and surface pressure have also been computed. The stability analysis has been performed successfully.

2.1 Solution for the wormhole shape function

A static, spherically symmetric wormhole is described by the line element

$$ds^2 = e^{\nu(r)}dt^2 - \frac{dr^2}{1 - \frac{b(r)}{r}} - r^2(d\theta^2 + sin^2\theta d\phi^2). \quad (1)$$

Here the radial metric potential $b(r)$ denotes the shape function as it represents the shape of the wormhole and the temporal metric potential $v(r)$ is the redshift function of the wormhole, which basically gives a measure of the redshift due to the loss in energy when a particle escapes the strong gravitational field of the wormhole due to emission from it.

The modified field equations in LQC scenario for the wormhole metric (1) turn out to have the form

$$\frac{b'}{r^2} = 8\pi\rho\left(1 - \frac{\rho}{\rho_c}\right), \tag{2}$$

$$\left(1 - \frac{b}{r}\right)\left(\frac{v'}{r} + \frac{1}{r^2}\right) - \frac{1}{r^2} = 8\pi\left(p - \frac{\rho(2p+\rho)}{\rho_c}\right), \tag{3}$$

$$\left(1 - \frac{b}{r}\right)\left(v'' + v'^2 + \frac{v'}{r}\right) - \frac{b'-b}{2r}\left(v' + \frac{1}{r}\right)$$
$$= 8\pi\left(p - \frac{\rho(2p+\rho)}{\rho_c}\right). \tag{4}$$

Here, the matter source is a perfect fluid having stress–energy tensor of the form $T^\mu_\nu = diag(\rho, -p, -p, -p)$. The fluid obeys a linear EoS $p(r) = \mu\rho(r)$. The parameter $\rho_c$ is extremely important in LQC as it denotes the maximum or critical density, beyond which the energy density cannot rise further, thus preventing the formation of a curvature singularity due to diverging energy densities. However, the curvature singularity at the throat of the wormhole is not due to diverging energy densities but due to diverging tidal forces in the radial and tangential directions and hence it remains to be seen whether such a curvature singularity can be resolved in the framework of LQC, giving rise to a stable traversable wormhole. As we seen in the RHS of Eqs. (2)–(4), the additional terms quadratic in stress energy arise due to the effective UV corrections to the space-time geometry in the classical picture. The are accounted for in the matter sector to provide an effective matter description. It is worth noting that the effective pressures in the radial and tangential directions are identical, resulting in an inherent pressure isotropy as contrasted to models of braneworld gravity where the anisotropy is generated from the extra dimensional contribution [41].

We have applied the equations for homogenous LQC to the spherically symmetric spacetime. There is an issue with the covariance in this class of models with consideration of local physical degrees of freedom and it was first shown by Bojowald and Brahma [49] that the covariance breaks on extending such models beyond a background treatment. The possible reason behind this lies in the non-Riemannian nature of the spacetime structures involved in such a treatment. On considering possible generalizations of the spacetime structures, covariance may be considered in the sense of realizing an identical count of gauge transformations compared to the classical theory with the exception of slicing independence prevalent in Riemannian geometry [50,51]. The spacetime structure of quantum corrected black hole geometries were studied [52] but due to certain misinterpretations of the quantum corrected phase space [53,54] and asymptotic [55] behaviour, some inconsistencies were found with the treatment [56]. A possible solution to this may involve field redefined metric components arising from certain generators of modified hypersurface deformations leading to the applicability of line elements in specific spacetime regions [57,58]. The most general covariant theory considering spherical symmetry have been derived at a canonical level [59]. The modified gravitational behaviour of symmetry reduced LQC models lack a covariant modified spherically symmetric solution [60]. A generalized form of covariance described by non-Riemannian geometry could be helpful. Our wormhole model constructed in the LQC setup is an elementary one and this is one of the main limitations that we hope to address in recent future.

The temporal metric potential is assumed to be given by the Kuchowicz function [61]

$$e^{v(r)} = e^{Br^2 + 2\ln C}, \tag{5}$$

where the constant parameter $B$ has dimension of inverse length squared while the parameter $C$ is a dimensionless constant. The reason behind the choice of the Kuchowicz potential as the redshift function has been stated in the discussion section.

The energy density of the matter inside the wormhole can be found making use of the redshift function and the considered linear EoS of the constituent matter

$$\rho(r) = C_1 e^{-\frac{(\mu+1)Br^2}{2\mu}}, \tag{6}$$

where $C_1$ denotes a constant of integration.

Making use of the expressions for the quantities available at hand in the modified EFE, the shape function can be obtained and is found to be given by

$$b(r) = \frac{6re^{-2Br^2}}{B\rho_c(\mu-1)(3\mu-1)(2Br^2+1)}\left(-\frac{8\pi C_1\mu^2\rho_c(\mu-1)}{3}\right.$$
$$\times e^{\frac{Br^2(3\mu-1)}{2\mu}} + \left(4\mu C_1^2\pi(2\mu+1)e^{\frac{Br^2(\mu-1)}{\mu}}\right.$$
$$\left.\left.+\rho_c\left(\left(Br^2+\frac{1}{2}\right)e^{2Br^2}+\frac{C_2}{2}\right)B(\mu-1)\right)\left(\mu-\frac{1}{3}\right)\right). \tag{7}$$

$C_2$ is another integration constant that may be determined from the junction conditions.

The shape function is plotted along the radial distance in Fig. 1 and it turns out to represent the shape of the wormhole quite well. The desired properties of the shape function to successfully describe a wormhole are also satisfied. At the throat radius $r_0 = 0.5$ km, the shape function has an identical value and the ratio $\frac{b(r)}{r}$ is well maintained to be less than unity

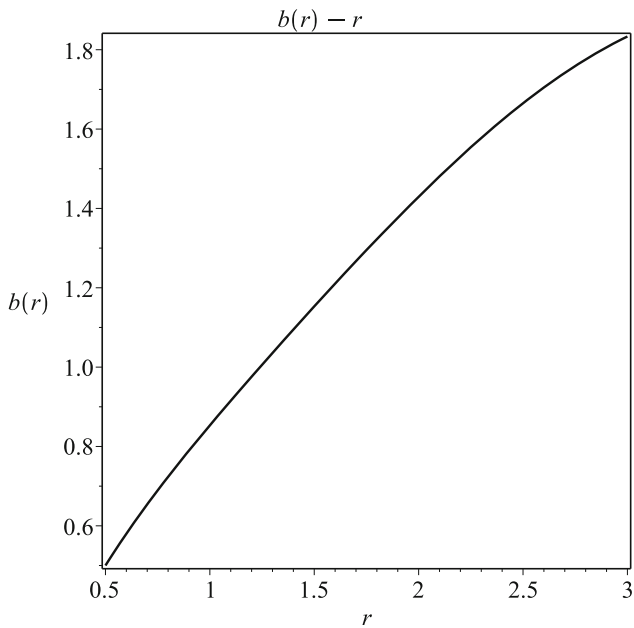

**Fig. 1** Variation of the shape function with respect to $r$

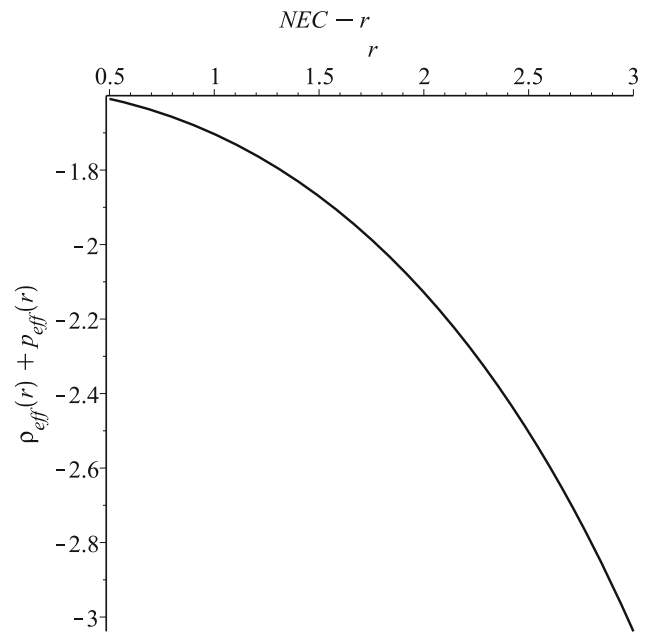

**Fig. 2** Variation of the NEC with respect to $r$

for all radial distances within the wormhole surface greater than the throat radius.

### 2.2 Validity of NEC

The geometrical modifications arising due to LQC can be effectively expressed as modifications in the matter sector, replacing the energy–momentum tensor of the perfect fluid matter source by an effective energy-momentum tensor of the form $T_\nu^{\mu(eff)} = diag(\rho^{eff}, -p^{eff}, -p^{eff}, -p^{eff})$, the time component of which is the effective energy density expressed as

$$\rho^{eff} = 8\pi \left( \rho \left( 1 - \frac{\rho}{\rho_c} \right) \right), \tag{8}$$

and the isotropic spatial components turn out to have the form

$$p^{eff} = 8\pi \left( p + \frac{\rho \left( p - \frac{\rho}{2} \right)}{\frac{\rho_c}{2}} \right). \tag{9}$$

Summing up the effective energy density and effective pressure

$$\rho_{eff} + p_{eff} = -\frac{16 C_1 (\mu + 1) \pi}{\rho_c} \left( C_1 e^{-\frac{(\mu+1)Br^2}{2\mu}} - \frac{\rho_c}{2} \right)$$
$$\times e^{-\frac{(\mu+1)Br^2}{2\mu}}. \tag{10}$$

We represent a plot of the variation of the summed up effective energy density and effective pressure along the radial expanse of the wormhole. As we see from Fig. 2, the sum is always a negative quantity within the wormhole and hence the NEC is effectively violated, although we find

$\mu > -1$ from the junction conditions as we shall see in the following subsection. So, we can say that it is the effective EoS parameter $\mu_{eff}$ arising from UV corrections that violates NEC as $\mu_{eff} < -1$.

### 2.3 The junction conditions

The spacetime exterior to the wormhole surface is a vacuum and can be described by the Schwarzschild line element which has the well known form

$$ds^2 = \left( 1 - \frac{2M}{r} \right) dt^2 - \left( 1 - \frac{2M}{r} \right)^{-1} dr^2$$
$$- r^2 (d\theta^2 + \sin^2 \theta d\phi^2), \tag{11}$$

where $M$ is the total mass of the wormhole.

The presence of matter at the wormhole surface leads to an extrinsic discontinuity resulting in a non-zero surface energy density and surface pressure. The wormhole surface behaves like a junction between the interior of the wormhole and the exterior Schwarzschild space-time in order to make the wormhole space-time geodesically complete. Thus, the junction conditions due to Israel and Darmois [62,63] are applicable at the wormhole surface resulting in continuity of the metric potential across it. Though, it does not always guarantee a continuity of the derivative of the metric potential across the surface. We compute the surface density and surface pressure using the junction conditions.

The intrinsic surface stress energy tensor is found to have the form $S_{ij} = diag(\Sigma, -P, -P, -P)$ that can be derived from the Lanczos equation [64–67].

In the most general form, $S_{ij}$ is defined using the Lanczos equation

$$S^i_j = -\frac{1}{8\pi}(\kappa^i_j - \delta^i_j \kappa^k_k), \quad (12)$$

where discontinuity in the extrinsic curvature across the surface is given by

$$\kappa_{ij} = \kappa^+_{ij} - \kappa^-_{ij}, \quad (13)$$

such that $+$ and $-$ implies the space-times interior and exterior to the wormhole surface. The second fundamental form can be obtained from the relation

$$\kappa^\pm_{ij} = -n^\pm_\nu \left[ \frac{\partial^2 X_\nu}{\partial \xi^i \partial \xi^j} + \Gamma^\nu_{\alpha\beta} \frac{\partial X^\alpha}{\partial \xi^i} \frac{\partial X^\beta}{\partial \xi^j} \right]|_S, \quad (14)$$

where $n^\pm_\nu$ denotes the normal vectors of unit magnitude defined as

$$n^\pm_\nu = \pm \left| g^{\alpha\beta} \frac{\partial f}{\partial X^\alpha} \frac{\partial f}{\partial X^\beta} \right|^{-\frac{1}{2}} \frac{\partial f}{\partial X^\nu}. \quad (15)$$

We consider $n^\nu n_\nu = 1$, while the intrinsic coordinate at the surface of the wormhole is represented by $\xi^i$ and satisfies the parametric equation $f(x^\alpha(\xi^i)) = 0$.

The surface energy density can be computed to have the form

$$\Sigma = -\frac{1}{2\pi R}\left[\sqrt{e^\lambda}\right]^+_- = \frac{1}{2\pi R}\left(\sqrt{1 - \frac{2M}{R}} - \sqrt{1 - \frac{6Re^{-2BR^2}}{B\rho_c HFG}\left(-\frac{8\pi C_1 \mu^2 \rho_c He^{\frac{BR^2 F}{2\mu}}}{3} + \left(4\mu C_1^2 \pi(2\mu+1)e^{\frac{BR^2 H}{\mu}} + \rho_c\left(\frac{G}{2}\right)e^{2BR^2} + \frac{C_2}{2}\right)BH\right)\frac{F}{3}}\right). \quad (16)$$

The surface pressure turns out to be given by

$$\mathcal{P} = \frac{1}{16\pi R}\left[\left(\frac{2f + f'R}{\sqrt{f}}\right)\right]^+_- = \frac{6e^{-2BR^2}}{\pi G^2 FBR^3 H\rho_c}\left(\frac{HG^2 e^{2BR^2}}{4}\left(MR - \frac{R^2}{2} - \frac{M}{2}\right)\frac{B\rho_c F}{3}\right.$$
$$\times \sqrt{\frac{1}{e^{2BR^2}B\rho_c HFG}\left(16e^{\frac{3BR^2}{2}}\pi\mu^2 C_1 \rho_c He^{\frac{BR^2}{2\mu}} - F\left(BC_2\rho_c He^{\frac{BR^2}{\mu}} + 8\left(2\mu+1\right)\pi C_1^2 e^{BR^2}\mu\right)\right)e^{-\frac{BR^2}{\mu}}}$$
$$+\frac{M'R^2}{3}\left(-\frac{4C_1 H}{3}\left(R^3(\mu+1)B^2 - 2BR^2\mu + \frac{(5\mu+1)BR}{2} - \mu\right)\mu\rho_c\pi e^{\frac{BR^2 F}{2\mu}} + \left(8\left(R^3(\mu+1)B^2 - BR^2\mu\right.\right.\right.$$
$$\left.\left.\left.+\frac{BR(3\mu+1)}{2} - \frac{\mu}{2}\right)C_1^2\left(\mu+\frac{1}{2}\right)\pi e^{\frac{BR^2 H}{\mu}} + HC_2 B\rho_c\left(B^2 R^3 - \frac{BR^2}{2} + BR - \frac{1}{4}\right)\right)\frac{F}{3}\right)\right)$$
$$\times \frac{1}{\sqrt{\frac{1}{e^{2BR^2}B\rho_c HFG}\left(16e^{\frac{3BR^2}{2}}\pi\mu^2 C_1\rho_c He^{\frac{BR^2}{2\mu}} - F\left(BC_2\rho_c He^{\frac{BR^2}{\mu}} + 16\left(\mu+\frac{1}{2}\right)\pi C_1^2 e^{BR^2}\mu\right)\right)e^{-\frac{BR^2}{\mu}}}} \frac{1}{M'}, \quad (17)$$

where $G = (2BR^2+1)$, $F = (3\mu-1)$, $H = (\mu-1)$, $M' = \sqrt{1-\frac{2M}{R}}$.

The wormhole space-time being static, the surface density and pressure shall vanish at the surface [31,41], the vanishing surface density giving the boundary condition

$$b(r)|_{r=R} = 2M. \quad (18)$$

The matching conditions to obtain the other unknown model parameters also appear from the junction conditions, where $g_{tt}|_{int} = g_{tt}|_{ext}$ and $\frac{\partial g_{tt}}{\partial r}|_{int} = \frac{\partial g_{tt}}{\partial r}|_{ext}$ at the surface of the wormhole $r = R$. So, we have three conditions in all.

We choose physically relevant values of the model parameters $B = 0.006$ km$^{-2}$, $\rho_c = 0.41$ m$^4$ and $M = 2.496\,M_\odot$ and making use of the boundary and matching conditions we obtain the unknown model parameters as $\mu = -0.9$, $C_1 = 0.4756683923$ and $C_2 = 0.150492837$. These value have been used to construct all the plots in this paper. As we obtain $\mu > -1$, the SEC is violated by physical matter but not the NEC. The NEC is however violated by the effective matter as the quadratic corrections make $\rho_{eff} + p_{eff} < 0$, implying $\mu_{eff} < -1$.

2.4 Tidal acceleration

The tidal acceleration experienced by the traveller at the throat of the wormhole must have both its radial and tangential components restricted to a reasonable value, which is usually considered to be the acceleration due to gravity on the earth. This shall ensure that the traveller crosses the wormhole throat safely and also we can obtain an upper limit on the velocity of the traveller while traversing the throat from the tangential acceleration.

The tidal acceleration along the radial direction is expressed by computing the $|R_{rtrt}|$ component of the Riemann tensor, which for the wormhole metric turns out to have the form

$$|R_{rtrt}| = \left|\left(1-\frac{b}{r}\right)\left[\frac{v''}{2}+\frac{v'^2}{4}-\frac{b'r-b}{2r(r-b)}\cdot\frac{v'}{2}\right]\right| \leq g_{earth}. \quad (19)$$

The condition is satisfied by our wormhole model.

The tidal acceleration along the tangential direction is found by computing two of the Riemann tensor components $|R_{\theta t\theta t}|$ and $|R_{\theta r\theta r}|$ and has the form

$$\gamma^2|R_{\theta t\theta t}|+\gamma^2 v^2|R_{\theta r\theta r}|=\left|\frac{\gamma^2}{2r^2}\left[v^2\left(b'-\frac{b}{r}\right)+(r-b)v'\right]\right|$$
$$\leq g_{earth}, \quad (20)$$

where, the $\gamma = \frac{1}{\sqrt{1-v^2}}$ represents the Lorentz factor, $v$ being the velocity with which the traveller traverses the wormhole throat. It seems reasonable to approximate that the velocity of the traveller at the throat is of the order much less than the velocity of light $v \ll 1$ implying a Lorentz factor $\gamma \equiv 1$. Making use of the assumed redshift function and the obtained shape function for the wormhole, the velocity of the traveller at the throat can be limited as

$$v \leq 0.099218371\sqrt{g_{earth}}, \quad (21)$$

which is a realistic limit that we obtain. The traversability of the wormhole can thus be ensured.

2.5 Linearized stability analysis

A linearized stability analysis is performed around the throat to ensure that our wormhole model is stable at the throat and remains traversable. For doing so, we consider the throat radius to be a proper time function $r_0 = x(\tau)$. This consideration gives the surface density and surface pressure, having the form

$$\Sigma = -\frac{1}{2\pi x}\sqrt{f(x)+\dot{x}^2}, \quad (22)$$

and

$$\mathcal{P} = \frac{1}{8\pi}\frac{f'(x)}{\sqrt{f(x)}}-\frac{\sigma}{2}, \quad (23)$$

where the function $f(x) = 1 - \frac{2M}{x}$, the parameter $M$ representing the wormhole mass.

The equation of motion can be obtained making use of the energy–momentum conservation as

$$\dot{x}^2 + V(x) = 0, \quad (24)$$

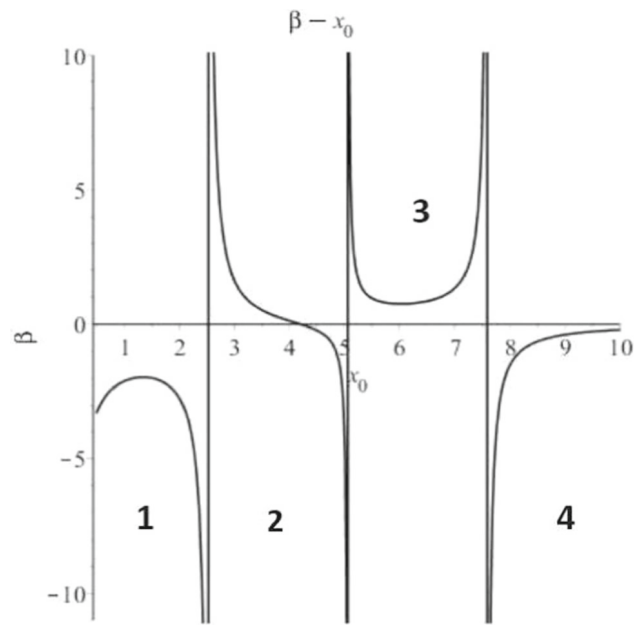

**Fig. 3** Plot of $\beta$ vs $x_0$

where, the effective potential $V(x)$ is constructed from the surface energy density, having the form

$$V(x) = f(x) - [2\pi x \Sigma(x)]^2. \quad (25)$$

A linearization is considered around the static solution $x_0$ which we assume for the equation of motion given by Eq. (24).

On expanding the constructed potential up to second order around the assumed solution $x_0$ using Taylor series, one can get

$$V(x) = V(x_0) - V'(x_0)(x-x_0) + \frac{1}{2}V''(x_0)(x-x_0)^2$$
$$+ O[(x-x_0)^3], \quad (26)$$

where prime implies derivative with respect to x.

For stability at the throat, the constructed effective potential must have a minimum at the throat which demands $V'(x_0) = 0$ and $V''(X_0) > 0$. The parameter $\beta = \frac{\delta \mathcal{P}}{\delta \Sigma}$ is introduced, in terms of which we shall express the condition for minima of the potential involving its second derivative as an inequality. The second derivative of the potential with respect to $x$ can be expressed in terms of the newly introduced parameter $\beta$ as

$$V''(x) = f''(x) - 8\pi^2[(\Sigma+2\mathcal{P})^2 + \Sigma(\Sigma+\mathcal{P})(1+2\beta)]. \quad (27)$$

This provides us with the stability condition at the throat in terms of $\beta$ as

$$\beta < \frac{\frac{f''(x_0)}{8\pi^2}-(\Sigma+2\mathcal{P})^2-2\Sigma(\Sigma+\mathcal{P})}{4\Sigma(\Sigma+\mathcal{P})}. \quad (28)$$

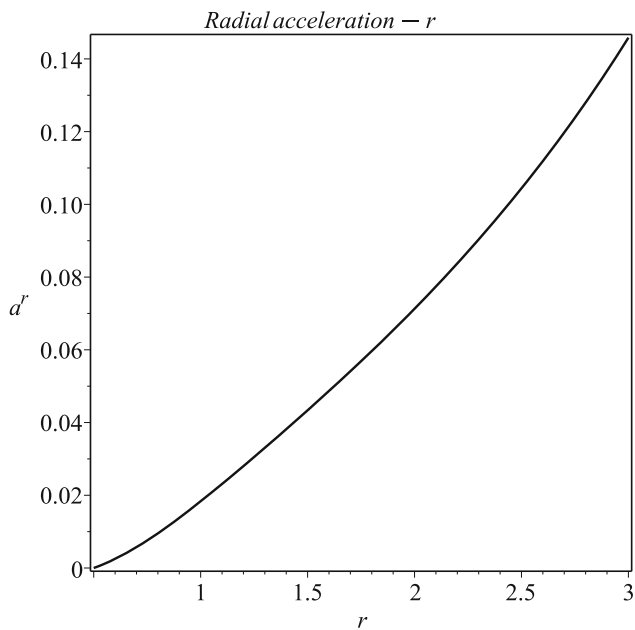

**Fig. 4** Plot of radial component of acceleration with respect to $r$

Using the relations for $\Sigma$ and $\mathcal{P}$, Eq. (28) can be written in the simplified form as

$$\beta < \frac{x_0^2 (f_0')^2 - 2x_0^2 f_0'' f_0}{4 f_0 (x_0 f_0' - 2 f_0)} - \frac{1}{2}. \tag{29}$$

For the wormhole we have constructed in the LQC scenario, the parameter $\beta$ turns out to have the value

$$\beta = \frac{10(2\pi - 1)m^2 + 3(-4\pi + 3)mr - 2r^2}{8\pi r(-r + 2m)(-r + 3m)}. \tag{30}$$

We have plotted the variation of the parameter $\beta$ along $x_0$ in Fig. 3. From the stability condition obtained using the minima of the constructed potential, the stable regions of the wormhole have been marked as regions 1–4 in the figure.

## 2.6 Acceleration and nature of the wormhole

It is interesting to compute the radial component of the four-acceleration for a static observer just outside the wormhole as if this component is positive, it implies that the wormhole features an attractive geometry implying that an outward directed radial acceleration is required on the static observer in order to stop being pulled into the wormhole. Likewise, if the radial component of the four acceleration is a negative quantity, the wormhole geometry is a repulsive one, implying the necessity of an inward directed radial acceleration on the static observer to prevent being pushed away from the wormhole.

A test particle initially at rest has the geodesic equation in the radial direction given by

$$\frac{d^2 r}{dt^\tau} = -\Gamma_{tt}^r \left(\frac{dt}{d\tau}\right)^2 = -a^r, \tag{31}$$

where $a^r$ is the radial 4-acceleration.

The four-velocity of a static observer near the wormhole is

$$U^\mu = \frac{dx^\mu}{d\tau} = (e^{-\frac{v(r)}{2}}, 0, 0, 0), \tag{32}$$

where $\tau$ denotes proper time as in the previous subsection.

Alternatively, the four-acceleration $a^\mu$ can be computed from the four-velocity as $a^\mu = U^\mu_{;\nu} U^\nu$, where the radial component of the 4-acceleration is expressed in terms of the metric potentials as

$$a^r = \frac{v'}{2}\left(1 - \frac{b(r)}{r}\right). \tag{33}$$

The radial component of the 4-acceleration for a static observer for the LQC constructed wormhole is given as

$$a^r = Br\left(1 - \frac{2e^{-2Br^2}}{B\rho_c HFG}\left(2\mu C_1^2 \pi (2\mu + 1)e^{\frac{Br^2 H}{\mu}}\rho_c \right.\right.$$
$$\left.\left. \times \left(Ge^{2Br^2} + C_2\right) BHF - 8\pi C_1 \mu^2 \rho_c H e^{\frac{Br^2 F}{2\mu}}\right)\right) v^2. \tag{34}$$

The variation of the radial 4-acceleration across the wormhole has been plotted in Fig. 4. The radial acceleration turns to be positive for all values of $r$ implying that the wormhole constructed in the LQC scenario features an attractive geometry, requiring an outward directed radial acceleration on the static observer to prevent from being pulled into the wormhole.

## 3 Discussions and conclusion

In this paper we have attempted to construct a traversable wormhole in the UV corrected framework of LQC. The classical EFE are modified by effective quadratic corrections in stress energy in the LQC scenario in an attempt to apply the central effects of Loop quantum gravity. Even the effective scenario is more fundamental in understanding the spacetime geometry and the strong curvature singularities appearing in GR can be averted without ambiguities arising from quantum mechanical interpretations. Solving the modified EFE for a static, spherically symmetric metric describing a wormhole spacetime, where the redshift function is assumed to be given by the Kuchowicz metric function which is well behaved in the vicinity of the wormhole and has been used to construct traversable wormhole on the RSII braneworld [41]. It is also found to work well in case of regular objects involving large space-time curvatures [68,69]. In relativistic wormhole solutions with modifications to the source sector involving exotic matter components, it often becomes impossible to obtain analytical solutions unless a constant redshift function is assumed [29,31]. This problem does not arise in the UV corrected scenario, where we consider a radially varying redshift function. The matter distribution at the wormhole throat is assumed to obey a linear EoS given by $p = \mu\rho$. From the modified field equations, the shape function of the wormhole is obtained. As expected, there is a dependence on the EoS parameter and the critical density, besides the other model parameters. On plotting the variation of the shape function with the radial distance, the plot turns out to represent the shape of the wormhole quite well, where the values of the model parameters are used as obtained from the junction conditions for generating the plot.

A model independent kinematical constraint on black hole bounce implying shell bounce in an untrapped region (either inside inner horizon or outside outer horizon) has been developed [70]. Extensions of the Oppenheimer–Snyder collapse in the form of black-to-white hole bounce have also been studied by matching the exterior static geometry with a spatially close FRW interior characterized by a bounce. A consistent model of black-to-white hole bounce applying the techniques of LQC has been constructed. So, it is established that LQC corrections can be used to model black-white hole bounces consistently. However, it may still be of some interest to study the possibility of existence of static traversable Lorentzian wormholes in LQC framework.

Causality should not be violated in a consistent wormhole even if the wormhole harbors traversability [71,72]. A key inconsistency or instability in models of traversable wormholes may appear from the apparent violation of causality owing to the possibilities of faster-than-light travel or travelling backwards in time. This may depend on what type of matter finds relevance in opening up the wormhole throat and the stability of the wormhole throat. However, the type of static Lorentzian wormhole that we have obtained in the LQC setup is "long" wormhole which implies a lesser travelling time in the ambient space surrounding the wormhole structure than through it and does not facilitate the formation of closed timelike curves. Moreover, the upper limit on the velocity of the traveller trying to traverse the wormhole experiencing tangential tidal forces within the desired limit as obtained from our analysis is considerably smaller than the velocity of light. Possible causality violation may lead to instabilities both at the classical and quantum level. A linearized stability analysis performed on the wormhole throat with the effective potential formalism indicates the stability of the throat. Also, most importantly we have not required any exotic matter violating the Null Energy condition to sustain an open wormhole throat due to the effective quantum corrections appearing from LQC. So, there is no real possibility of faster-than-light travel or backward time travel thus ensuring that causality is not violated. However, a better understanding of the spacetime structure in spherically symmetric setup in LQC may help us establish this issue more comprehensively in the recent future.

The validity of the NEC is checked for the effective matter distribution. Although the physical matter at the wormhole throat does not violate the NEC, *the NEC is effectively violated due to the UV correction terms arising in the modified EFE which are quadratic in stress energy. So, we do not require exotic matter to construct the traversable wormhole.* The obtained value of the EoS parameter from the junction conditions imply that the physical matter at the throat must violate the SEC in order to make the wormhole traversable. As discussed earlier, such matter can cause the universe to accelerate. This leads to the interesting result that *dark energy can be harnessed into a wormhole in the framework of LQC, such that the EoS parameter inside the throat of the wormhole $\mu > -1$ but $\mu_{eff} < -1$ due to the LQC corrections.* No additional ingredient is required in the energy budget of the universe to construct a traversable wormhole. Moreover, as dark energy is the dominant component of the energy budget at present, so any wormhole that is formed in the present epoch does not require its mass to be minimized as in the standard relativistic context. More importantly, the quantum instabilities [76] arising from physical matter violating the NEC can be avoided. For the observed acceleration of the universe, it is indeed required that $-1.61 < \mu < -0.78$ [73–75], but the pathologies associated with such violation of NEC using matter sector [76] are difficult to tackle. So, a geometric modification is also of interest at the IR scale where $\mu_{eff} < -1$ without violation of NEC by physical matter, but the quadratic correction term will not remain significant at these scales due to low energy densities and some alternative mechanism must prevail.

The tidal acceleration obtained at the throat is within desirable limits both in the radial and tangential directions and the consequent upper limit obtained on the velocity of the traveller traversing the wormhole throat is a realistic one. So, any traveller trying to use the wormhole as a shortcut for space-time travel does not get ripped apart at the throat of the wormhole due to infinitely large tidal forces in either the radial or tangential direction. One can think that despite the matter at the throat not violating the NEC (as $\mu > -1$), *the tidal forces and hence the Weyl curvature tensor do not diverge due to the effect of LQC which prevents the tidal force from increasing beyond a certain limit, thus resolving the singularity.* This ensures the traversability of the wormhole. However, we can say that LQC is more effective in resolving the strong curvature singularities that arise from diverging energy densities, as it can do so without violating any of the energy conditions. The reason behind this may be that for a diverging spacetime curvature due to infinitely large energy densities like the initial singularity of the universe or one at the center of a Schwarzschild black hole, the Weyl curvature vanishes rather than diverging, contrary to the behaviour at the wormhole throat. So, these singularities can be completely resolved in a LQC context (which has an inherent pressure isotropy in the UV corrected EFE) without matter violating any of the energy conditions. However, to resolve the diverging Weyl curvature at the wormhole throat without violating any of the energy conditions, the presence of an inherent anisotropy in the UV-corrected EFE may play a significant role as indicated by braneworld (which has an inherent pressure anisotropy due to contribution of the bulk Weyl tensor projected on the brane) wormholes, which can be constructed from matter obeying all the energy conditions [41].

Performing a linear stability analysis check, we can say that the wormhole would not collapse at the throat due to development of any instability and remains traversable. In order to perform the check, an effective potential formalism is applied, where the radius of the wormhole throat is assumed to be a function of the proper time and an effective potential is constructed from the surface density of the wormhole, which is in turn obtained from the junction conditions at the wormhole surface. The equation of motion in terms of this effective potential can be obtained from the conservation of stress–energy. A parameter $\beta$ is introduced involving the surface pressure and surface density. The stability condition essentially represents a minima of the potential, obtained from the vanishing of its first derivative and the second derivative being positive at the throat. The condition for this minima is expressed as an inequality in terms of the parameter $\beta$ that is plotted to obtain the regions of stability as denoted. The radial acceleration remaining positive for radial distances within the wormhole, it can be said to feature an attractive geometry.

Although a wormhole has not been detected yet till date, but there is the possibility of detecting one in the recent future. A number of suggestions have been proposed in literature to detect one [77–82]. This can be done by studying any unexplained effect on the orbital motion of stars near black holes that can harbor wormholes [83]. Also, micro-lensing effects of wormholes have been suggested to resemble gamma ray bursts [84]. Emission of radiation pulses remains another interesting possibility [85]. Wormhole that are constructed from phantom matter violating the NEC with a particular EoS can be distinguished from black holes via the process of quasinormal ringing [86]. If a wormhole is indeed detected in the recent future, then detailed observational studies can also throw light on the actual nature of the UV corrected gravity due to the large space-time curvatures involved. For the time being we can conclude by stating that *LQC not only resolves the curvature singularities due to diverging energy densities, leading to non-singular bouncing black holes* [87] *but also due to diverging tidal forces, leading to traversable wormholes without any NEC violating exotic matter which may result in quantum instabilities*.

**Acknowledgements** MK is thankful to the Inter-University Centre for Astronomy and Astrophysics (IUCAA), Pune, India for providing the Visiting Associateship under which a part of this work was carried out. RS is thankful to the Govt. of West Bengal for financial support through SVMCM scheme. SG is thankful to the Directorate of Legal Metrology under the Department of Consumer Affairs, West Bengal for their support.

**Data Availability Statement** This manuscript has no associated data or the data will not be deposited. [Authors' comment: This is a theoretical work and no data has been used to arrive at any of the results.]